\documentclass[preprint,5p,times,twocolumn]{elsarticle}
\usepackage{graphicx,color,epsfig,rotate,amssymb}

\journal{Solid State Communications}

\begin{document}

\begin{frontmatter}

\title{Phase transitions in a spinless, extended Falicov-Kimball model on the triangular lattice}

\author{Umesh K. Yadav, T. Maitra and Ishwar Singh}

\address{Department of Physics, Indian Institute of Technology Roorkee, Roorkee- 247667, Uttarakhand, India}

\begin{abstract}
A numerical diagonalization technique with canonical Monte-Carlo simulation algorithm is
used to study the phase transitions from low temperature (ordered) phase to
high temperature (disordered) phase of spinless Falicov-Kimball model on a
triangular lattice with correlated hopping ($t^{\prime}$). It is observed that
the low temperature ordered phases (i.e. regular, bounded and segregated)
persist up to a finite critical temperature ($T_{c}$). In addition, we observe
that the critical temperature decreases with increasing the correlated hopping
in regular and bounded phases whereas it increases in the segregated
phase. Single and multi peak patterns seen in the temperature dependence of
specific heat ($C_v$) and charge susceptibility ($\chi$) for different values
of parameters like on-site Coulomb correlation strength ($U$), correlated
hopping ($t^{\prime}$) and filling of localized electrons ($n_{f}$) are also
discussed.
\end{abstract}


\end{frontmatter}

\section{Introduction}
The study of correlated systems like transition metal dichalcogenides~\cite{aebi,qian2,cava},
cobaltates~\cite{qian}, $GdI_{2}$~\cite{gd1,gd2} and its doped variant  $GdI_{2}H_{x}$~\cite{gd3}
have attracted considerable attention recently as they exhibit a number of
remarkable cooperative phenomena like valence and metal-insulator transition, charge and magnetic
order, excitonic instability and possible non-Fermi liquid states~\cite{gd2,castro}. It has been suggested
recently that these correlated systems may very well be described by different extensions of the
Falicov-Kimball model (FKM) on a triangular lattice~\cite{gd2,umesh1,umesh2,umesh3,umesh4}. These systems are
geometrically frustrated. The frustration gives rise to a large degeneracy at low temperatures and competing
ground states close by in energy. A consequence of this is a fairly complex ground state
magnetic phase diagram~\cite{gd3} and the presence of soft local modes
strongly coupled to the conduction electrons~\cite{gd2}.

The usual FKM~\cite{fkm} was proposed to study the semiconductor-metal
transition in $SmB_{6}$ and other correlated electron systems. The FKM considers only
the kinetic energy of the itinerant electrons and the local Coulomb interaction
between the itinerant and localized electrons. The effective interactions in the FKM are
mediated by the itinerant electrons~\cite{kennedy}. FKM has been applied to different types
of lattices (bipartite/non-bipartite) and in different dimensions. While FKM on bipartite lattices
is studied extensively~(see Ref.~\cite{free} and references therein), results on non-bipartite lattices
are rare~(see Ref.~\cite{umesh1,maska}).

Recently several new ground states were observed for an extended FKM on
a triangular lattice where the effect of correlated hopping $t^{\prime}$, was
taken into account~\cite{umesh1}. The correlated hopping term is influenced by the number of localized
electrons (denoted by `$f$'-electrons) present on the neighboring sites as explained below. For example,
in some rare earth compounds (especially the mixed-valence compounds), the rare earth ions with two
different ionic configurations $f^{n}$ and $f^{n-1}$ have different ionic radii: $f^{n-1}$ configuration
has less screening of nuclear charge compared to $f^{n}$ configuration. The itinerant electrons
(denoted by `$d$'-electrons) in ions with $f^{n-1}$ configuration feel more attraction due to the
nucleus and hence get contracted. This leads to increased localization of those orbitals. Hence
the $d$-orbital overlap between nearest neighbors depends on local $f$-electron occupation of
neighboring ions, resulting in a correlated hopping of $d$-electrons. Such a correlated hopping
term also appears in the first principles calculation~\cite{hirsch} of the tight binding Hamiltonian and is
usually neglected in Hubbard type models~\cite{umesh1}. Its significance in superconductivity
has been emphasized already~\cite{hirsch2,bulk} and in the context of FKM, it has been considered by several
authors~\cite{wojt,shv,farkov}.

Therefore, we consider such an extended FK Hamiltonian on a triangular lattice
\begin{eqnarray}
{H} =-\sum_{\langle ij\rangle}(t_{ij}+\mu\delta_{ij})d^{\dagger}_{i}d_{j}
+E_f \sum_{i} f^{\dagger}_{i}f_{i}
\nonumber \\
+U \sum_{i}f^{\dagger}_{i}f_{i}d^{\dagger}_{i}d_{i}
+\sum_{\langle ij\rangle}t^{\prime}_{ij}(f^{\dagger}_{i}f_{i}+f^{\dagger}_{j}f_{j})d^{\dagger}_{i}d_{j}.
\end{eqnarray}
\noindent where $d^{\dagger}_{i},\, d_{i}\, (f^{\dagger}_{i}, f_{i}$) are, respectively, the creation
and annihilation operators for $d\, (f)$-electrons at site $i$ and $\mu$ is the chemical potential.
The first term in Eq.(1) is nearest-neighbor hopping of $d$-electrons on a
triangular lattice. The second term represents the dispersionless energy level $E_{f}$ of $f$-electrons while the
third term is the on-site Coulomb repulsion between $d$ and $f$-electrons. The last term is the correlated hopping
discussed above.

A comprehensive study of the ground state phase diagram of Hamiltonian Eq.(1) is reported elsewhere~\cite{umesh1}.
With correlated hopping ($t^{\prime}\,\in [-1,1]$) two different phases are observed at $\frac{1}{4}$-filling
of localized electrons namely (i) ordered and (ii) segregated phases for $U=1$,
$3$ and $5$. At $\frac{1}{2}$-filling of localized electrons (i) bounded (ii) axial striped and (iii) segregated
phases for $U=1$ and (i) striped and (ii) segregated phases for $U=5$ are reported. Some of these charge ordered
phases are also observed experimentally in the above mentioned systems. In this work, we extend our study to finite
temperatures and study the half filled situation for the Hamiltonian Eq.(1) and observe the stability of
these phases and the nature of the phase transitions at finite temperature.

%%%%%%%%%%%%%%%%%%%%%%%%%%%%%%%%%%%%%%
\section{Methodology}
%%%%%%%%%%%%%%%%%%%%%%%%%%%%%%%%%%%%%%
In the Hamiltonian Eq.(1), the local $f$-electron occupation number $\hat{n}_{f,i}=f^{\dagger}_{i}f_{i}$
commutes with $H$ and $\omega_{i}=f^{\dagger}_{i}f_{i}$ is a good quantum number
taking values either $1$ or $0$. This local gauge symmetry also
implies~\cite{elitzur} that interband excitonic averages of the type
$\langle d_{i}^{\dagger}f_{j} \rangle$ are identically zero at any
finite temperature (i.e., an absence of homogeneous mixed valence)
and the $f$-electron level remains dispersionless. Using this local conservation the Hamiltonian
may be written as,
\begin{equation}
H=\sum_{\langle ij\rangle}h_{ij}(\omega)d^{+}_{i}d_{j}+E_f\sum_{i}\omega_{i}
\end{equation}
\noindent where $h_{ij}(\omega)=[-t_{ij}+t^{\prime}_{ij}
(\omega_{i}+\omega_{j})]+(U\omega_{i}-\mu)
\delta_{ij}$.

The value of $\mu$ is chosen such that $N_{f}+N_{d}=N$ (half-filled case),
where $N_{f},\, N_{d}$ are the total number of $f$ and $d$-electrons and $N=L^{2}~(L=12)$
is the number of sites. For a lattice of $N$ sites, the Hamiltonian $H$ is now an $N\times N$
matrix for a fixed configuration $\omega$. We set the value of hopping
$t_{\langle ij\rangle}$=1. We choose a particular value of $N_{f}$ ($0 \le N_{f} \le N$),
and a configuration $\omega=\{\omega_1,\omega_2, \dots , \omega_N \}$ for $N_f$. Choosing values
for $t^{\prime}$ and $U$, the eigenvalues $\lambda_{i}$ of $H(\omega)$, are calculated by numerical
diagonalization with the periodic boundary conditions (PBC). Average value of the physical quantities
is calculated by Monte Carlo sampling method reported elsewhere~\cite{umesh1}.

The partition function is written as
\begin{equation}
{Z}=\prod_{i}\left(\sum_{{\omega_i=0,1}}\,
Tr\, e^{-\beta H(\{\omega_i\})} \right)
\end{equation}
\noindent where the trace is taken over the $d$-electrons, and $\beta=\frac{1}{kT}$.
The trace is calculated from the eigenvalues $\lambda_{i}\, (i=1\cdots N)$ of the
matrix $h$ (first term in $H$ (Eq.(2)).

\noindent The partition function can, therefore, be recast in the form
\begin{eqnarray}
{Z}=\prod_{i}(\sum_{\omega_i=0,1}e^{-\beta E_{f}\omega_{i}})\,\prod_{j}^{N}
(1+e^{-\beta[\lambda_{j}(\omega)-\mu]})
\end{eqnarray}

The corresponding total internal energy is
\begin{eqnarray}
{E(\omega)}= \sum_{i}{\frac{\lambda_{i}(\omega)}{[e^{\lambda_{i}(\omega)/kT}+1]}}+\,N_{f}E_{f}
\end{eqnarray}

The thermodynamic quantities are calculated by the averages with
the statistical weights,
\begin{eqnarray}
{A(\omega)}=\frac{1}{Z}e^{-\beta E(\omega)},
\end{eqnarray}
of localized f-electron configurations $\omega$, respectively.

In the simulations, one starts with random configuration (of $f-$electrons) at high
temperature and then the temperature is ramped down slowly until a stable ground state
is obtained at very low temperature. Then this ground state configuration is chosen as the
initial configuration for a particular finite temperature. For a fixed finite temperature a
large number of Monte Carlo steps are carried out to make the system reach the equilibrium.
We have observed that around $1000$ Monte Carlo steps are sufficient for this. Then the ensemble
average of the thermodynamic quantity of interest is calculated on next ($\sim 4000$) Monte Carlo
steps. Now an important question arises whether the system has reached equilibrium or not.
To ensure this we rerun the simulation starting from two different initial configurations;
firstly from a fully ordered configuration of $f-$electrons and secondly from random configuration
of $f-$ electrons. After a certain number of Monte Carlo steps (around 1000 steps) the internal
energies for both the cases approach same value. This implies that the system has reached equilibrium.
A similar approach was earlier employed in the study of phase transitions in FKM on a square
lattice~\cite{maska,vries}.

We have calculated temperature dependence of the charge susceptibility $\chi$ and the specific heat $C_{v}$.
The $\chi$  and $C_v$ are calculated by the fluctuation-dissipation theorem (FDT). The $\chi$ is related through
the FDT to the first order variance of the density-density correlation function~\cite{maska}, given as,
\begin{eqnarray}
{\chi}=\frac{1}{kT}(\langle {\textrm{g}}_{n=1}^2 \rangle - {\langle {\textrm{g}}_{n=1} \rangle}^{2}),
\end{eqnarray}
where $\langle \dots \rangle$ represent average over different configurations. The $\textrm{g}_n$ is
the density-density correlation function defined as,
\begin{eqnarray}
{{\textrm{g}}_n}={\frac{1}{N}}\,{\sum_{i}^N}\,{\sum_{\tau_{1},\,{\tau_{2}=\pm{n}}}\,
\omega(\textbf{r}_i)\,\omega(\textbf{r}_{i}+ \tau_{1}\hat{a}_{x}+\tau_{2}\hat{a}_{y}}),
\end{eqnarray}
where $\omega(\textbf{r}_{i})=\omega_{i}$, $\hat{a}_{x}$ and $\hat{a}_{y}$ are the unitary
vectors along the x and y directions respectively (here we assumed lattice constant as unity).

The specific heat ($C_v$) is related through the FDT to the internal energy and defined as,
\begin{eqnarray}
{C_{v}}=\frac{1}{N}\,\frac{1}{kT^2}(\langle E^2 \rangle - {\langle E \rangle}^{2})
\end{eqnarray}

The specific heat and charge susceptibility are calculated by different approaches. The specific
heat is calculated from eigenvalues of the $d-$electron Hamiltonian. The $d-$electron spectrum
changes with temperature. Correspondingly total internal energy of $d-$electrons changes with
temperature. This change in total internal energy of the system with temperature is nothing but
temperature dependence of the specific heat. It has been seen that in a particular temperature
range a few eigenvalues in the lower part of $d-$band increase with increase in temperature
(while other eigenvalues are remaining unaffected). This manifest in the increase of $d-$electron
specific heat around that temperature. The charge susceptibility is calculated from density-density
correlation function ($g_{n}$). The $g_{n}$ depends on $f-$electron configuration only.

In addition, we have calculated the thermal average (or the ensemble average) of the $f-$electron
occupation ($\omega_{i}(T)$) on each site ($i=1~to~N$).
The $\omega_{i}(T)$ on each site (i) is given as,
\begin{eqnarray}
{\omega_{i}(T)}=\langle \omega_{i} \rangle
               =\frac{\sum_{j}\,{\omega_{i}\,e^{-\beta E(\{\omega_{j}\})}}}{\sum_{j}e^{-\beta E(\{\omega_{j}\})}},
\end{eqnarray}
where, summation is taken over the different Monte Carlo simulation steps at a finite temperature.
We have chosen a random configuration of $f-$electrons initially. At $T=0$ an ordered ground state configuration
of $f-$ electrons is obtained. At a finite temperature $T$, this ordered configuration is chosen as initial
configuration. The ensemble average of the $f-$electron occupation, $\omega_{i}(T)$, is calculated for
$\sim 4000$ Monte Carlo steps, after the system reached equilibrium.

Phase transition from low temperature ordered phase to high temperature homogeneous
or disordered phase is characterized by the temperature dependence of thermal average $f-$ electron occupation,
$\omega_{i}(T)$. In case of `first-order' phase transition, discontinuous change in $\omega_{i}(T)$ from
ordered phase to homogeneous phase with temperature is found. In `second-order' phase transition continuous change in
$\omega_{i}(T)$ from ordered phase to homogeneous phase with temperature is seen. At $T_c$ (which corresponds
to last peak in $C_v$ and $\chi$) homogeneous configuration of $f-$electrons is found.

%%%%%%%%%%%%%%%%%%%%%%%%%%%%%%%%%%%%%%%
\section{Results and discussion}
%%%%%%%%%%%%%%%%%%%%%%%%%%%%%%%%%%%%%%%%%\

%%%%%%%%%%%%%%%%%%%%%%%%%%%%%%%%%%%%%%%%%%%%%%%%%%%%%%%%%%%%
\begin{figure*}[ht]
\begin{center}
\includegraphics[trim = 0.5mm 0.5mm 0.5mm 0.5mm, clip,width=10.50cm]{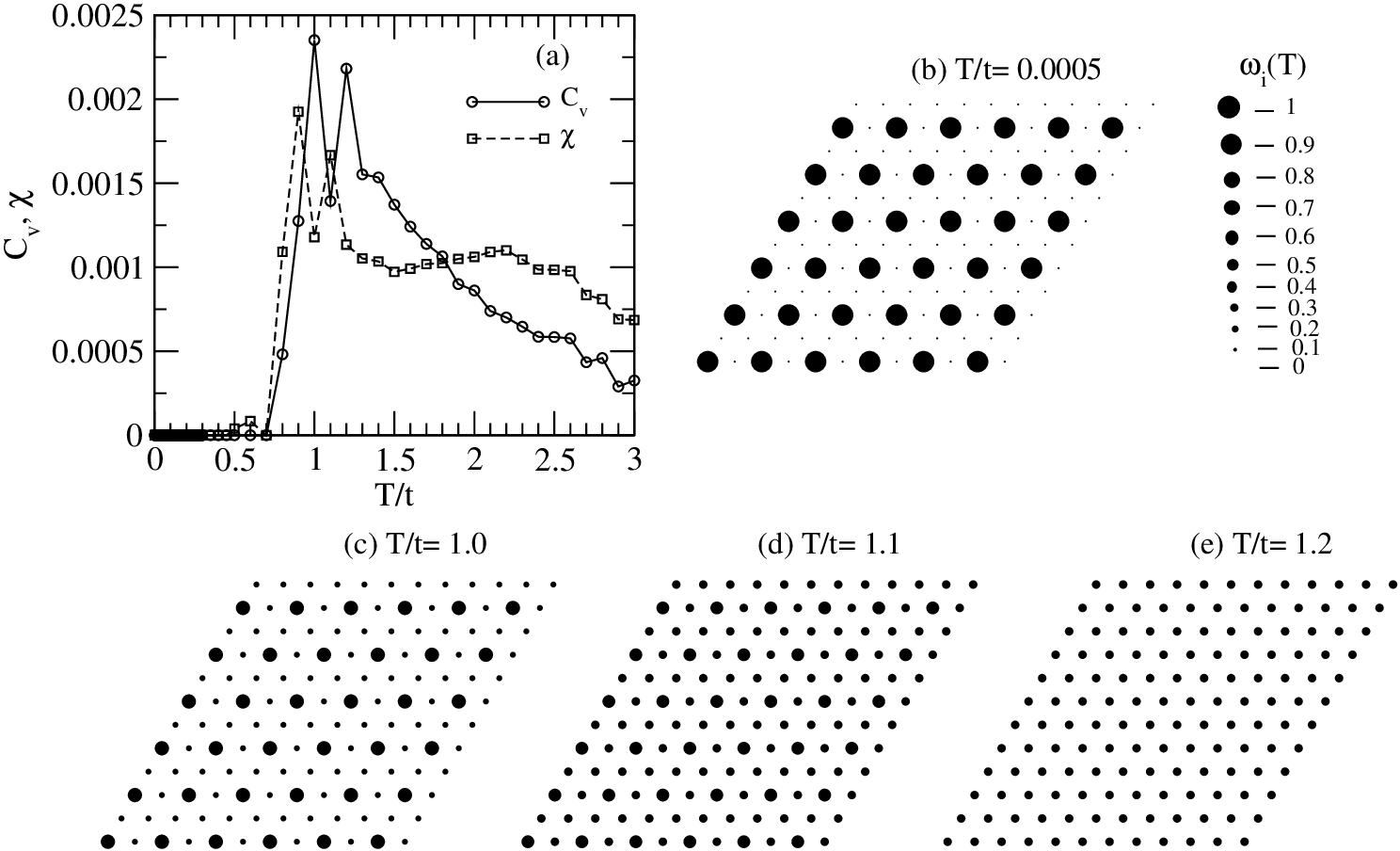}
\caption{The specific heat $C_v$ and charge susceptibility $\chi$ (a) and average f-electron
occupation $\omega_{i}(T)$ on each site ((b) to (e)) for different temperatures at U=1,
$t^{\prime}$=-1.0 and $n_{f}=\frac{1}{4}$. Strength of the filled circles are shown
in right side of upper panel.}
\label{Fig1}
\end{center}
\end{figure*}
%%%%%%%%%%%%%%%%%%%%%%%%%%%%%%%%%%%%%%%%%%%%%%%%%%%%%%%%%%%

At finite temperature, only a few exact calculation results are available on the bipartite (square) lattice
with~\cite{farkov_arxiv} and without~\cite{maska} correlated hopping. Without correlated hopping (usual FKM) and for
half-filled case, it was observed that for all finite Coulomb correlation $U>0$ the ground state configuration is
checkerboard type that persist up to a critical temperature $T_{c}(U)$. At $T_c$ the system undergoes a phase transition
from ordered phase to the homogeneous phase. The phase transition is of first order for $U < 1$ and of second order
for $U>1$ ~\cite{maska,farkov_arxiv}. With correlated hopping $t^{\prime}$ and $U=0.5$, low temperature phase
is an ordered one and mainly, is of type checkerboard, axial striped and segregated phase. For $U=0.5$, and
for $-1.0 \le t^{\prime} \le 0.55$ the phase transition is of the first order and for $t^{\prime}> 0.55$,
phase transition is of the second order. In the regime of $t^{\prime}$, where phase is of the checkerboard type
critical temperature reduces with increasing $t^{\prime}$ and in the regime where phase is of the type of axial
striped and segregated, the critical temperature enhances with increase in the $t^{\prime}$.

Perturbative results for large $U$ indicate that to order $1/U$, the FKM can be mapped onto
an Ising antiferromagnet (AFM) in a magnetic field,
$H_{eff}=\sum_{\langle ij \rangle} \frac{t^2}{4U}s_{i}s_{j}+\frac{1}{2}(\mu+E_{f})\sum_{i}s_{i}
+{\rm constant\,terms}$, where $s_{i}=2\omega_{i}-1,\, s_{i} = -1,1$.
The Ising AFM state on a triangular lattice is frustrated and leads to large
degeneracies at low temperature. It turns out that this frustration is lifted~\cite{gruber}
in higher order perturbation in $1/U$. It is therefore quite intriguing that one would expect
the effects of frustration to bear on the finite temperature states as on-site Coulomb correlation ($U$)
between $d-$ and $f-$ electrons and the chemical potential ($\mu$) are varied.

There is hardly any exact result for the FKM on the non-bipartite lattice at
finite temperature~\cite{maska_tri}. Therefore, it is quite interesting to observe
the temperature induced phase transitions in the FKM on a triangular lattice. We have
studied the finite temperature induced phase transitions as a function of a range of values
of $U$ and fillings (i) $n_{f}=N_{f}/N=\frac{1}{4}$  and (ii) $n_{f}=\frac{1}{2}$\,($N_d$ is
constrained to $N-N_{f}$) of $f-$electrons using the method outlined in the previous section.
We look at the effect of correlated hopping of $d$-electrons on phase transitions for several
values of $t^{\prime} \in [-1,1]$ at a fixed $U$.

We have studied variation of specific heat (C$_v$) and charge susceptibility
($\chi$) with temperature for some fixed values of U and $t^\prime$. We find
that in the cases C$_v$ and $\chi$ remain zero up to certain temperature (say
T$_1$) and then on further increasing the temperature two peak like structure
appear in both C$_v$ and $\chi$ for nonzero values of $t^\prime$ (One observes
a single peak in case of $t^\prime=0$). The second peak in case of $t^\prime$ nonzero
(and the single peak in case of $t^\prime=0$) corresponds to the temperature
(T$_c$) at which a phase transition is seen to occur from the ordered to the
homogeneous phase as discussed below in terms of the temperature dependence of
the order parameter. We have also studied the thermal average of the f-electron
occupation at each site ($\omega(T)$ = $<\omega_i>$) given by equation (10).

%%%%%%%%%%%%%%%%%%%%%%%%%%%%%%%%%%%%%%%
\subsection{$n_{f}=\frac{1}{4}$ case:}
%%%%%%%%%%%%%%%%%%%%%%%%%%%%%%%%%%%%%%%

%%%%%%%%%%%%%%%%%%%%%%%%%%%%%%%%%%%%%%%%%%%%%%%
\begin{figure*}[ht]
\begin{center}
\includegraphics[trim=0.5mm 0.5mm 0.5mm 0.5mm,clip,width=10.50cm]{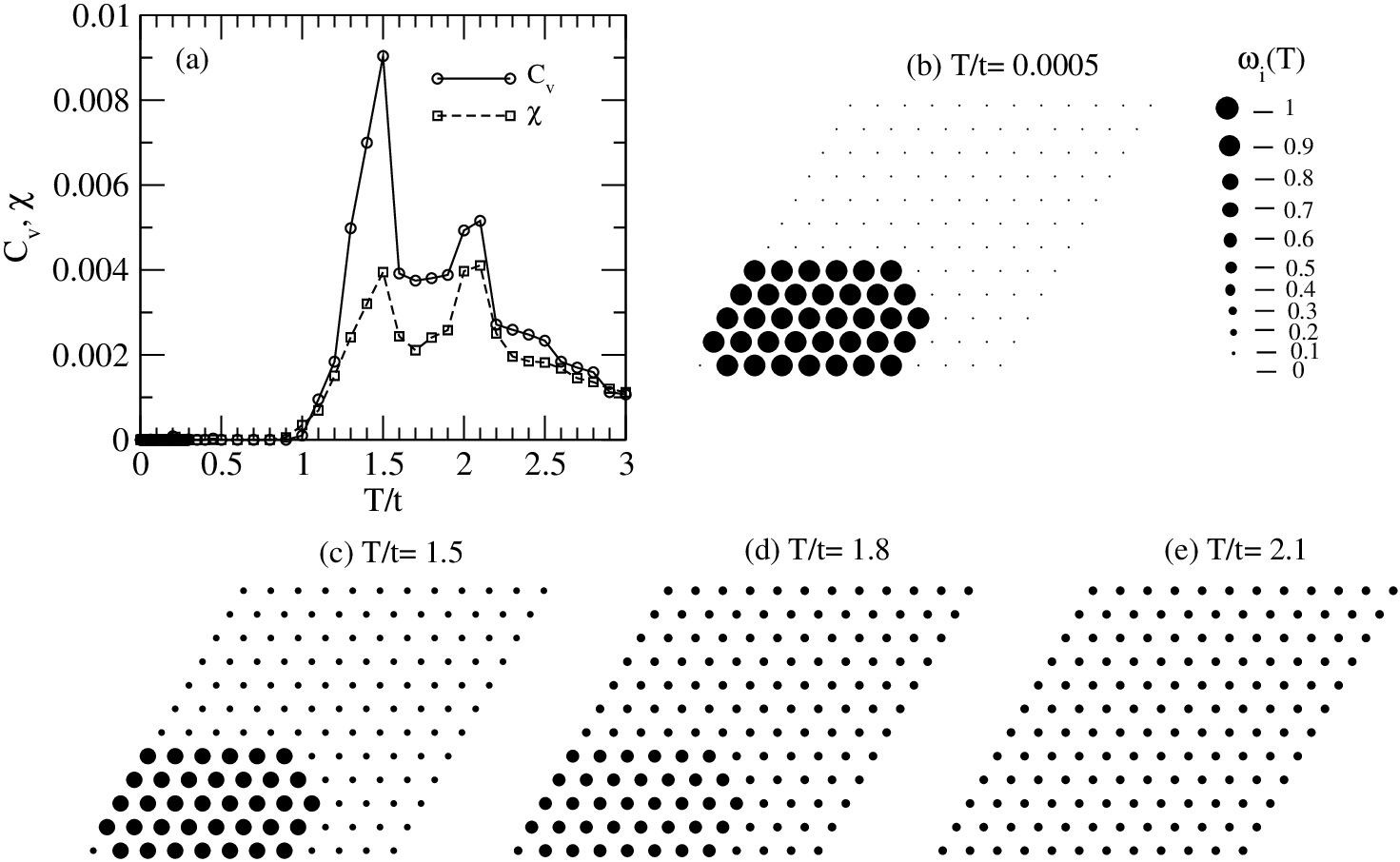}
\caption{The specific heat $C_v$ and charge susceptibility $\chi$ (a) and average
f-electron occupation $\omega_{i}(T)$ on each site ((b) to (e)) for different
temperatures at U=1, $t^{\prime}$=1.0 and $n_{f}=\frac{1}{4}$. Strength of the
filled circles are shown in right side of upper panel.}
\label{Fig2}
\end{center}
\end{figure*}
%%%%%%%%%%%%%%%%%%%%%%%%%%%%%%%%%%%%%%%%%%%%%%%%%%%%%%%%%%%%%%

Two different phases namely ordered and segregated phase are observed for correlated hopping
~$t^{\prime}\,\in [-1,1]$ and $U=1$~\cite{umesh1}. An ordered ground state configuration of $f-$electrons is observed
for $t^{\prime}$=-1.0 to $\sim$ 0.35 and $U=1$. In Fig.$1$(a) we present the variation of specific heat ($C_v$)
and charge susceptibility ($\chi$) with temperature for $t^{\prime}$=-1.0 and $U=1$. The $C_v$
and $\chi$ have values zero up to temperature $T_{1} = 0.7$ and then two peak like structure appear with second
peak in $C_v$ at $T_{c} = 1.2$. The thermal average of the $f-$electron occupation at each site ($\omega_{i}(T)$)
given by Eq.$10$ at $t^{\prime}$=-1.0 and $U=1$ for different temperatures is shown in Fig.$1$ (b), (c) (d) and (e).
The ground state configuration of $f-$electrons is ordered: $\omega_{i}=1$ on one-fourth of the sites and $\omega_{i}=0$
on the rest. This behavior of the $f-$ electrons configuration persists up to a finite temperature $T_1 \sim$ $0.7$.
Above this temperature, the average $f-$electron occupation at each site changes continuously between $1$ and $0$
(i.e. In between temperature $T_{1}$ and $T_{c}$, $\omega_{i}(T)$ decreases from $1$ at sites with $\omega_{i}=1$
below $T_{1}$ and increases from $0$ at sites with $\omega_{i}=0$ (See the discussion below in terms of the order parameter).
At the critical temperature $T_{c} (\sim$ 1.2), a homogeneous phase with the $f-$electrons occupying equally all the
sites is observed which is also the temperature at which second peak appears in
$C_v$. The rise in $C_v$ above $T_1$ is understandable because when order parameter starts decreasing from $1$
the entropy starts increasing and thus contributing to the rise in $C_v$. The appearance of the first peak is
still not clearly understood though. We believe that it could be an artefact of finite lattice size as also seen
in case of bipartite lattices (Fig.$6$ of Ref. $25$). Even though in square lattice the first peak is seen to get
suppressed with a lattice size $16\times 16$, in the triangular lattice case the lattice size at which this suppression will happen could probably be much higher.

The ground state $f-$electrons configuration for $t^{\prime} > 0.35$ and $U=1$, is segregated phase.
The above mentioned quantities ($C_v$, $\chi$ and $\omega_{i}(T)$) are calculated for correlated hopping $t^{\prime}$=1.0
and $U=1$ (see Fig.$2$). Two peak structure is observed in $C_v$ and $\chi$. Temperature dependence of $\omega_{i}(T)$
suggests second order phase transition from segregated phase to homogeneous phase. In this region the critical temperature
$T_c$, where phase transition occurs, increases with increasing the correlated hopping $t^{\prime}$.

To observe the effect of $t^{\prime}$ on phase transition for $U>1$, we have calculated $C_v$,
$\chi$ and $\omega_{i}(T)$ for $U=5$ and $t^{\prime} \in[-1,1]$ (not shown here). The phase transition occurs at
a higher value of critical temperature in comparison to that at $U=1$ for all calculated values of $t^{\prime}$.
Due to large on-site Coulomb interaction between $d-$ and $f-$electrons, large temperature is required to make the
$f-$electron occupation homogeneous.

%%%%%%%%%%%%%%%%%%%%%%%%%%%%%%%%%%%%%%%
\subsection{$n_{f}=\frac{1}{2}$ case:}
%%%%%%%%%%%%%%%%%%%%%%%%%%%%%%%%%%%%%%%

%%%%%%%%%%%%%%%%%%%%%%%%%%%%%%%%%%%%%%%%%%%%%%%
\begin{figure*}[ht]
\begin{center}
\includegraphics[trim=0.5mm 0.5mm 0.5mm 0.5mm,clip,width=10.50cm]{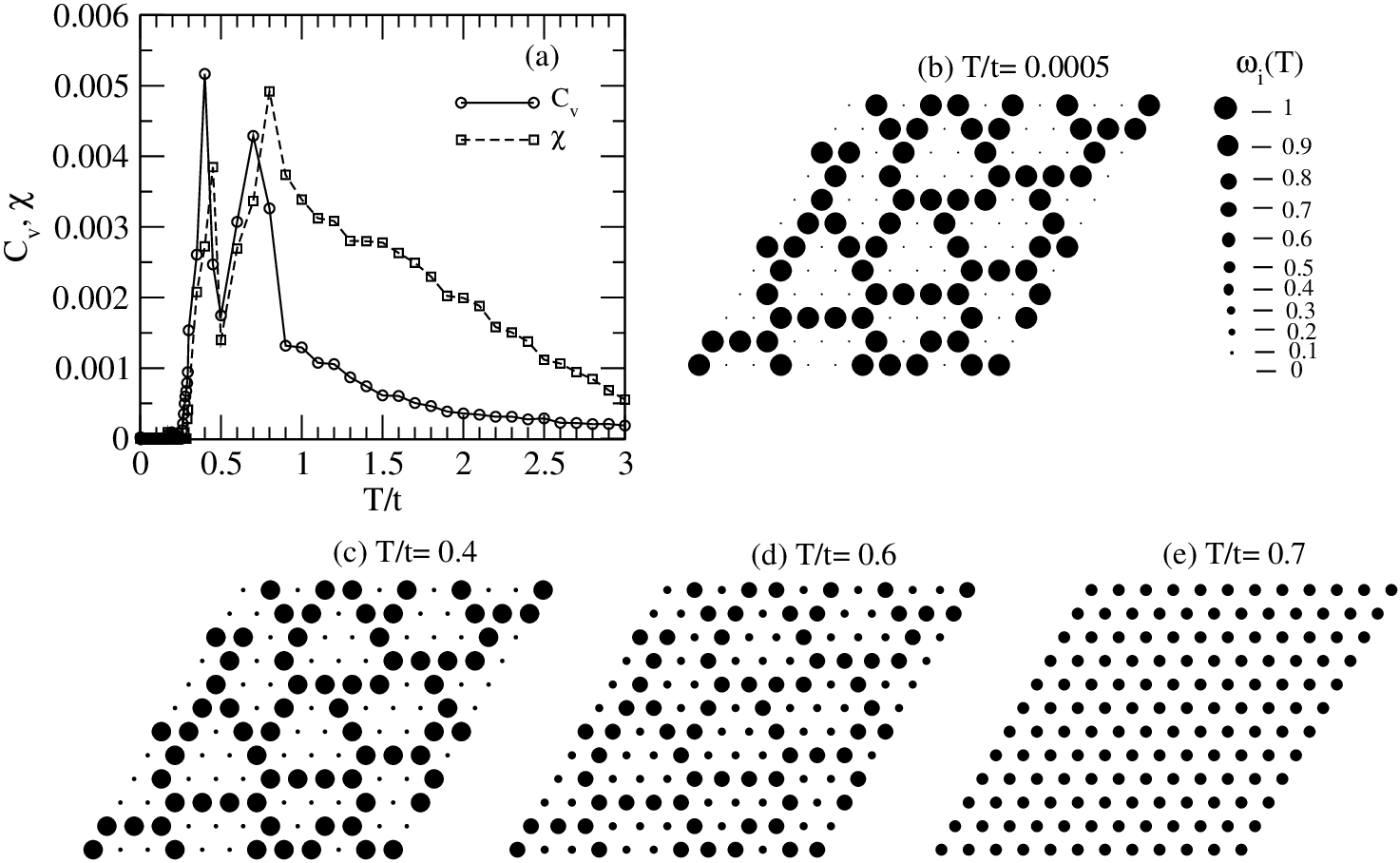}
\caption{The specific heat $C_v$ and charge susceptibility $\chi$ (a) and average f-electron occupation
$\omega_{i}(T)$ on each site ((b) to (e)) for different temperatures at U=1, $t^{\prime}$=-1.0 and $n_{f}=\frac{1}{2}$.
Strength of the filled circles are shown in right side of upper panel.}
\label{Fig3}
\end{center}
\end{figure*}
%%%%%%%%%%%%%%%%%%%%%%%%%%%%%%%%%%%%%%%%%%%%%%%%%%%%%%%%%%%%%%
%%%%%%%%%%%%%%%%%%%%%%%%%%%%%%%%%%%%%%%%%%%%%%%
\begin{figure*}[ht]
\begin{center}
\includegraphics[trim=0.5mm 0.5mm 0.5mm 0.5mm,clip,width=10.50cm]{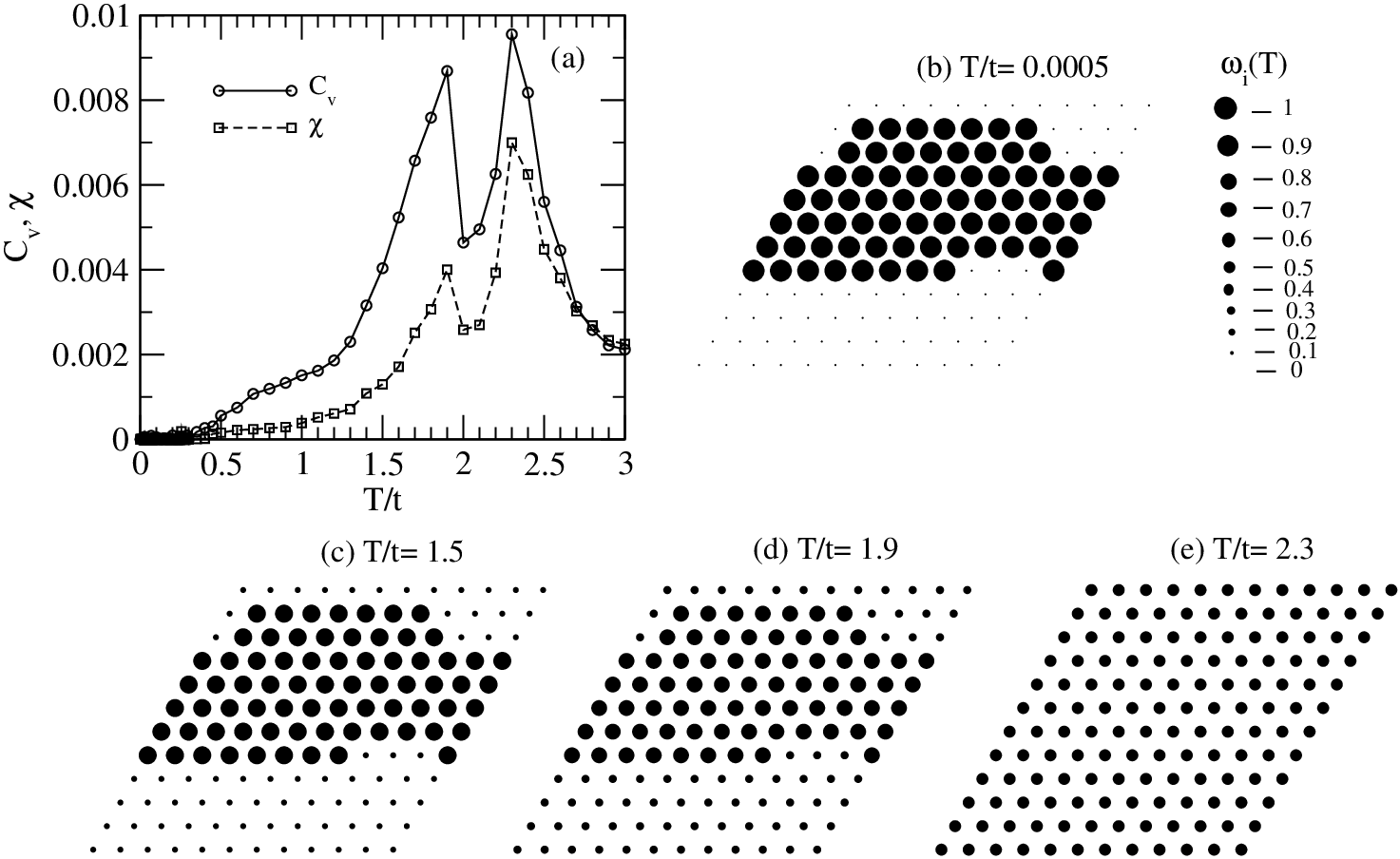}
\caption{The specific heat $C_v$ and charge susceptibility $\chi$ (a) and average f-electron
occupation $\omega_{i}(T)$ on each site ((b) to (e)) for different temperatures at U=1,
$t^{\prime}$=1.0 and $n_{f}=\frac{1}{2}$. Strength of the filled circles are shown in right side of upper panel.}
\label{Fig4}
\end{center}
\end{figure*}
%%%%%%%%%%%%%%%%%%%%%%%%%%%%%%%%%%%%%%%%%%%%%%%%%%%%%%%%%%%%%%

Three different ground state configurations namely bounded, axial striped and segregated phases
are observed~\cite{umesh1} at $U=1$ and $t^{\prime} \in [-1,1]$. In order to observe the stability
of these phases at finite temperatures and to see the effect of filling of $f$-electrons on $T_c$,
we have calculated $C_v$, $\chi$ and $\omega_{i}(T)$ at $U=1$ for $t^{\prime} \in [-1,1]$. Shown
in Figs.$3$ and $4$ are variation of $C_v$, $\chi$ and $\omega_{i}(T)$ for $U=1$ and for
$t^{\prime}$=$-1$ and $1$, respectively.

The ground state $f-$electrons configuration for $t^{\prime}$=-1.0 to $\sim$ 0.4 and $U=1$,
is bounded. At finite temperature, for all negative values of $t^{\prime}$, two peak structure
is observed in $C_v$ and $\chi$ (see Fig$3$(a) for $t^{\prime}$=-1.0). Corresponding variation
of $\omega_{i}(T)$ at different temperatures is shown in Figs.$3$(b), (c), (d) and (e). Temperature
dependence of $\omega_{i}(T)$ suggests second order phase transition from low temperature bounded
phase to high temperature homogeneous phase at $T_c$, which corresponds to the second peak in $C_v$ and $\chi$.

Segregated ground state $f-$electrons configuration is observed for $t^{\prime} \geq$ $0.6$ and $U=1$.
Variation of $C_v$, $\chi$ and $\omega_{i}(T)$ for $t^{\prime}$=1 and $U=1$ at different temperatures
is shown in Fig.$4$. Two-peak structure is observed in $C_v$ and $\chi$. Phase transition
of second order from low temperature segregated phase to high temperature homogeneous phase is
observed at $T_c$=2.3, which corresponds to the second peak in $C_v$ and $\chi$. The $T_c$
is very large in segregated phase in comparison to the other phases.

We have calculated $C_v$, $\chi$ and $\omega_{i}(T)$ for $U=5$ and $t^{\prime} \in [-1, 1]$
(not shown here). Multi-peaks are observed in $C_v$ and $\chi$. At all values of $t^{\prime}$
second order phase transition from low temperature ordered phase to high temperature disordered
phase about $T_c$ (much larger than at $U=1$), corresponding to the second peak in $C_v$ and $\chi$ is observed.

%%%%%%%%%%%%%%%%%%%%%%%%%%%%%%%%%%%%%%%%%%%%%%%
\begin{figure*}[ht]
\begin{center}
\includegraphics[trim=0.5mm 0.5mm 0.5mm 0.5mm,clip,width=8.50cm]{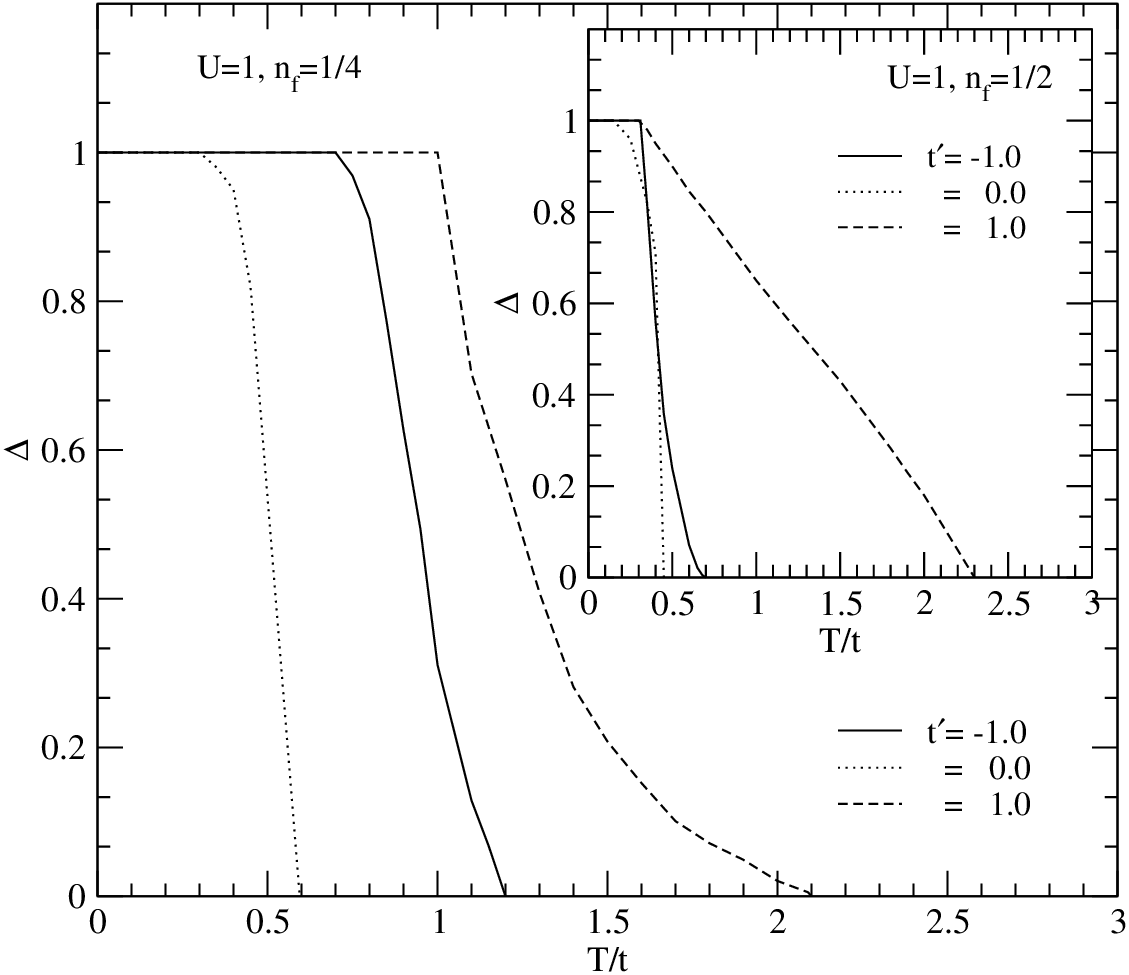}
\caption{Temperature dependence of order parameter $\Delta$ of the ordered phases for $U=1$, $n_{f}=\frac{1}{4}$
and $t^{\prime}$=$-1.0$, $0$ and $1.0$. The inset shows temperature dependence of $\Delta$ for
$U=1$, $n_{f}=\frac{1}{2}$ and $t^{\prime}$=$-1.0$, $0$ and $1.0$.}
\label{Fig5}
\end{center}
\end{figure*}
%%%%%%%%%%%%%%%%%%%%%%%%%%%%%%%%%%%%%%%%%%%%%%%%%%%%%%%%%%%%%%

In Fig.$5$ we have shown the temperature dependence of order parameter $\Delta$ ($\Delta=\langle{\omega_{A}}\rangle
-\langle{\omega_{B}}\rangle$, where $A$ and $B$ are occupied and unoccupied sites by $f-$electron at $T=0$ respectively
and $\langle \omega_{A}\rangle$ and $\langle \omega_{B}\rangle$ are thermal average of the $f-$electron occupation
(see Eqn. $10$) at site $A$ and $B$ for a particular temperature respectively) of the ordered phases for $U=1$ and
$t^{\prime}$=$-1$, $0$ and $1$ for $n_{f}=\frac{1}{4}$ and $n_{f}=\frac{1}{2}$ (shown in the inset).
Let us consider the case for $U=1, t^{\prime}=-1$ and $n_f=\frac{1}{4}$, $\Delta$ remains $1$ up
to $T_1=0.7$. It starts decreasing for $T_1>0.7$. In between $T_1>0.7$ and $T_{c}=1.2$ first peak in $C_v$ and $\chi$ is
seen at $T=1$.  At $T_c=1.2$, $\Delta$ vanishes and homogeneous configuration of $f$ electrons is found.
This temperature corresponds to the second peak in $C_v$ and $\chi$. As $\Delta$ decreases continuously with temperature
from a value $1$ to $0$, the phase transition can be categorized as a second order phase transition. For all the three
$t^{\prime}$ values (1, 0 and -1) we see similar behavior with the transition temperature being different in each case.
However, the behavior of the order parameter near the critical temperature is not like a typical mean field one
(The order parameter falls almost linearly). This kind of the behavior has been observed in the bipartite lattices as well
(Ref.$16$) where the authors attributed this effects to finite size of the lattice considered and the weak coupling limit.

In conclusion, we have studied the phase transitions in the spinless extended FKM on a triangular lattice
and found that the ground state ordered configuration persist up to a finite critical temperature ($T_c$).
Above $T_c$, homogeneous, disordered phase is observed. At $U=1$, $n_{f}=\frac{1}{4}$ and $\frac{1}{2}$, second
order phase transition from ordered phase to disordered phase is observed for values of $t^{\prime}= -1,~0$ and $1$.
At one-fourth filling of $f-$electrons critical temperature $T_c$, decreases in ordered phase and increases in
segregated phase with $t^{\prime}$. At half filling of $f$-electrons, segregated phase exists at higher values of
$t^{\prime}$ than at one-fourth filling. Correspondingly, $T_c$ decreases up to a higher value of
$t^{\prime}$ in half filling than one-fourth filling. The $T_c$, in ordered
and in bounded phase is larger and similar in segregated
phase for one fourth filling, in comparison to half filling. Second order
phase transition is found for all calculated values of $t^{\prime}$
and fillings of $f-$electrons at $U=5$. There is rarely any calculation available for
the spinless FKM on a triangular lattice at finite temperatures. Our results may motivate further studies of the
finite temperature properties of frustrated systems of recent interest like cobaltates, $GdI_{2}$, $NaTiO_{2}$ and $NaVO_{2}$.

%%%%%%%%%%%%%%%%%%%%%%%%%%%%%%%%%%%%%%%%%%%%%%%%%%%%%%%%%%%%%%
$Acknowledgments.$ UKY acknowledges CSIR, India for research fellowship.
%%%%%%%%%%%%%%%%%%%%%%%%%%%%%%%%%%%%%%%%%%%%%%%%%%%%%%%%%

%%%%%%%%%%%%%%%%%%%%%%%%%%%%%%%%%%%%%%%%%%%%%%%


\begin{thebibliography}{}
%%%%%%%%%%%%%%%%%%%%%%%%%%%%%%%%%%%%%%%%%%%%%%%%%%%
\bibitem{aebi} H. Cercellier, C. Monney, F. Clerc, C. Battaglia, L. Despont,  M. G. Garnier, H. Beck, and P. Aebi,
Phys. Rev. Lett. {\bf 99}, 146403 (2007).
\bibitem{qian2} E. Morosan, H. W. Zandbergen, B. S. Dennis, J. W. G. Bos,  Y. Onose, T. Klimczuk,
A. P. Ramirez, N. P. Ong, and R. J. Cava, Nature Phys. {\bf 2},
544 (2006); D. Qian, D. Hsieh, L. Wray, E. Morosan, N. L. Wang, Y. Xia, R. J. Cava, and M.Z. Hasan,
Phys. Rev. Lett. {\bf 98}, 117007 (2007); G. Li, W. Z. Hu, D. Qian, D. Hsieh, M. Z. Hasan, E. Morosan,
R. J. Cava, and N. L. Wang, Phys. Rev. Lett. {\bf 99}, 027404 (2007).
\bibitem{cava} K. E. Wagner, E. Morosan, Y. S. Hor, J. Tao, Y. Zhu, T. Sanders, T. M. McQueen, H. W. Zandbegen,
A. J. Williams, D. V. West, and R. J. Cava, Phys. Rev. B {\bf 78}, 104520 (2008).
\bibitem{qian} D. Qian, D. Hsieh, L. Wray, Y. D. Chuang, A. Fedorov, D. Wu, J. L. Luo,  N. L. Wang, L. Viciu,
R. J. Cava, and M. Z. Hasaan,
Phys. Rev. Lett. {\bf 96}, 216405 (2006).
\bibitem{gd1} C. Felser, K.. Ahn, R. K. Kremer, R. Seshadri, and A. Simon, Solid State Chem., {\bf 147}, 19 (1999).
\bibitem{gd2} A. Taraphder, L. Craco and M. Laad, Phys. Rev. Lett. {\bf 101}, 136410 (2008).
\bibitem{gd3} T. Maitra, A. Taraphder, A. N. Yaresko, and P.Fulde, Euro. Phys. Journal B {\bf 49}, 433 (2006).
\bibitem{castro} A. H. Castro Neto, Phys. Rev. Lett. {\bf 86}, 4383 (2001).
\bibitem{umesh1} Umesh K Yadav, T. Maitra, Ishwar Singh and A. Taraphder, J. Phys.: Condens. Matter {\bf 22}, 295602 (2010).
\bibitem{umesh2} Umesh K Yadav, T. Maitra, Ishwar Singh and A. Taraphder, EPL {\bf 93}, 47013 (2011).
\bibitem{umesh3} Umesh K Yadav, T. Maitra and Ishwar Singh, AIP Conf. Proc. {\bf 1349}, 109 (2011).
\bibitem{umesh4} Umesh K Yadav, T. Maitra and Ishwar Singh, Eur. Phys. J. B {\bf 84}, 365 (2011).
\bibitem{fkm} L. M. Falicov and J. C. Kimball, Phys. Rev. Lett. {\bf 22}, 997 (1969).
\bibitem{kennedy} T. Kennedy, Rev. Math. Phys. {\bf 6}, 901 (1994); T. Kennedy and E. H. Lieb, Physica A {\bf 138}, 320 (1986).
\bibitem{free} J. K. Freericks and V. Zlatic, Rev. Mod. Phys. {\bf 75}, 1333 (2003); P. S. Goldbaum,
J. Phys. A: Math. Gen. {\bf 36}, 2227 (2003).
\bibitem{maska} M. Maska and K. Czajka, Phys. Rev. B {\bf 74}, 035109 (2006).
\bibitem{hirsch} J. E. Hirsch, Physica A {\bf 158}, 326 (1989).
\bibitem{hirsch2} J. E. Hirsch, Phys. Rev. B {\bf 39}, 11515 (1989).
\bibitem{bulk} B. R. Bulka, Phys. Rev. B {\bf 57}, 10303 (1998).
\bibitem{wojt} J. Wojtkiewicz and R. Lemanski, Phys. Rev. B {\bf 64}, 233103 (2001).
\bibitem{shv} A. Shvaika, Phys. Rev. B {\bf 67}, 075101 (2003).
\bibitem{farkov} P. Farkasovsky and N. Hudakova, J. Phys. Cond-mat. {\bf 14}, 499 (2002).
\bibitem{elitzur} S. Elitzur, Phys. Rev. D {\bf 12}, 3978 (1975).
\bibitem{vries} Vries, Z. Phys. B {\bf 92}, 353 (1993).
\bibitem{farkov_arxiv}H. Cencarikova and P. Farkasovsky, arxiv:1007.2770 (2010).
\bibitem{gruber} C. Gruber, N. Macris, A. Messager and D. Ueltschi, J. Stat. Phys. {\bf 86}, 57 (1997).
\bibitem{maska_tri} K. Czajka and  M. Maska, Physica B {\bf 378-380}, 275-277 (2006).

\end{thebibliography}
\end{document}